\documentclass[fleqn,usenatbib]{mnras}

\usepackage{newtxtext,newtxmath}

\usepackage[T1]{fontenc}

\DeclareRobustCommand{\VAN}[3]{#2}
\let\VANthebibliography\thebibliography
\def\thebibliography{\DeclareRobustCommand{\VAN}[3]{##3}\VANthebibliography}

\usepackage{graphicx}	
\usepackage{amsmath}	
\usepackage{url}
\usepackage{xcolor}
\usepackage{soul}

\newcommand{\kms} {\TextOrMath{\,km s$^{-1}$}{\mathrm{\,km s}^{-1}}}
\newcommand{\msol}{\TextOrMath{\,M$_\odot$}{\mathrm{\,M}_\odot}} 

\newcommand{\rsol} {\TextOrMath{\,R$_\odot$}{\mathrm{\,R}_\odot}}
\newcommand{\about} {$\sim$}

\title[VFTS 243 in BPASS]{VFTS 243 as predicted by the BPASS fiducial models}

\author[H. F. Stevance et al.]{
H. F. Stevance$^{1,2}$\thanks{E-mail: hfstevance@gmail.com},
S. Ghodla$^{1}$,
S. Richards$^{1}$,
J. J. Eldridge$^{1}$,
M. M. Briel$^{1}$,
P. Tang$^{1}$,
\\
$^{1}$Department of Physics, The University of Auckland, Private Bag 92019, Auckland, New Zealand\\
$^{2}$Astrophysics Research Centre, School of Mathematics and Physics, Queen’s
University Belfast, N. Ireland, BT7 1NN, United Kingdom\\
}
\date{Accepted XXX. Received YYY; in original form ZZZ}

\pubyear{2022}

\begin{document}

\label{firstpage}
\pagerange{\pageref{firstpage}--\pageref{lastpage}}
\maketitle

\begin{abstract}
The recent discovery of an unambiguous quiescent BH and main sequence O star companion in VFTS 243 opens the door to new constraints on theoretical stellar evolution and population models looking to reproduce the progenitors of black hole - black hole binaries.
Here we show that the Binary Population and Spectral Synthesis fiducial models (BPASSv2.2.1) natively predict VFTS 243-like systems:
We find that VFTS 243 likely originated from a binary system in a \about 15 day orbit with primary mass ranging from 40 to 50 \msol\, and secondary star with initial mass 24--25\msol. 
Additionally we find that the death of the primary star must have resulted in a low energy explosion $E<10^{50}$ ergs.
With a uniform prior we find that the kick velocity of the new-born black hole was $\le$10 \kms.
The very low eccentricity reported for VFTS~243 and the subsequent conclusion by the authors that the SN kick must have been very small is in line with the peak in the posterior distribution between 0 and 5 \kms found from  our numerical simulations performed with a uniform prior.
Finally, the reduced Hobbs kick distribution commonly used in black hole population synthesis is strongly disfavoured. 
\end{abstract}

\begin{keywords}
stars: evolution -- stars: binaries: general -- stars: black holes --  stars: individual: VFTS 243
\end{keywords}



\section{Introduction}

The majority of massive stars (born with a mass greater than 8\msol) are present in binary systems \citep{sana2012,sana2014,moe17} and on exhausting their nuclear supply, they collapse under their gravitational pressure to form a compact remnant. If the collapsing star is sufficiently massive, this can create a binary system composed of a bright massive star in orbit around a black hole (BH). Such binary systems could be strong emitters of X-rays fueled by the accretion of matter from the star onto the BH companion \citep{Sakura_Sunyaev_1973}. Such a distinctive observational feature makes the identification of the binary system hosting the BH possible. 
Additionally, attempts have been made to identify Star+BH binaries in systems where the BH is quiescent (not X-ray loud) through spectroscopic studies. In recent years a number of such systems have been reported (LB-1 \citealt{liu2019}; HR 6819 \citealt{rivinius2020}; NGC 1850-BH1 \citealt{saracino2021}), only to be later suspected to harbour a stripped star as a BH impostor (\citealt{shenar2020,el-badry2020,bodensteiner2020,el-badry2021,stevance2022}) -- see \cite{bodensteiner2022} for a review.

The VFTS 243 system however seems to be the first undisputed discovery of a quiescent BH in a stellar binary.
Reported by \cite{shenar2022}, it is found to be an O7+BH in the 30 Doradus region. 
The very precise measurements of the orbital components (P = 10.4031 $\pm 0.0004$  days, \textit{e} = 0.017 $\pm 0.012$) and extensive observational data surrounding this object have the potential to provide improved observational constraints or clues on a number of bottlenecks in stellar evolution theory such as natal supernova (SN) kicks and the genealogies of the progenitor systems of BH+BH binaries. 
Detectable gravitational wave sources are, for the most part, BH-BH merger systems \citep{theligoscientificcollaboration2021c} but our collective understanding of their progenitor routes remains hazy \citep{mandel2021}.
This is in part due to the wide array of possible channels and the degeneracy between these routes when the only observations available are at the beginning (binary star systems) and the end (merger of compact objects) of the life of the systems. 
Being able to study binaries comprised of a star and a compact remnant is therefore essential to anchor our theoretical models to observations at a crucial intermediate stage in the life of gravitational wave progenitors.

In the supplementary material of the VFTS 243 BH discovery paper, \cite{shenar2022} present an exemplar system calculated with \textsc{mesa} \citep{Paxton2011, Paxton2013, Paxton2015, Paxton2018, Paxton2019} to show a possible progenitor route to the formation of VFTS 243-like objects. 
The flexibility of \textsc{mesa} allows observational parameters to be reproduced rather nicely (generally speaking and in the case of VFTS 243) but in order to truly understand the most likely genealogies of such systems population synthesis is required. 
The Binary Population And Spectral Synthesis (BPASS) project combines detailed binary stellar evolution (using a custom version of the Cambridge STARS code \citealt{eggleton1971,eldridge08}) with population synthesis based on observationally inferred binary distributions.
The fiducial models (BPASSv.2.2.1) were released in 2018 \citep{stanway2018} and have since been shown to reproduce a large number of observable phenomena, particularly in the realm of massive stars (e.g. \citealt{massey2021}). 
The key feature is that BPASS results are not tailored to fit specific observations -- they are pre-calculated models from a grid of initial parameters. 
This allows us to take a different approach when  comparing observational systems and theory: instead of creating a model to match the observations as best as possible, we can search through our existing predictions and find systems which are similar to those observed (e.g.  \citealt{xiao19,tang2020,stanway2020,stevance2020,stevance2021,byrne2021,briel2022,chrimes2022,ghodla2022,stevance2022}).

In this work we search the fiducial BPASS models for predictions of VFTS243-like systems and compare our genealogies to those of the exemplar  \textsc{mesa} model presented in \cite{shenar2022}.
In addition we investigate the natal kick velocities that best reproduce the orbital configuration of VFTS 243 and compare these to commonly used natal kick prescriptions for BH remnants. 
In Section \ref{sec:methods} we describe our models and numerical methods; in Section \ref{sec:results} we show our posterior kick velocities and best genealogies to VFTS 243 as predicted by BPASS. 
In Section \ref{sec:discussion} we discuss the differences between our channel and the \textsc{mesa} model in \cite{shenar2022}, as well as the implications of our kick posteriors. 
Finally we summarise and conclude in Section \ref{sec:conclusion}.


\section{Models and Numerical Methods}
\label{sec:methods}
We use the fiducial BPASSv2.2.1 models (described in \citealt{eldridge2017, stanway2018}): each population is born from an instantaneous starbust of 10$^6$ \msol\, at a single metallicity; in the present case we use Z=0.006 as it best corresponds to the average metallicity of the Large Magellanic Cloud (LMC -- \citealt{massey2021}).
Although a wide scatter of metallicities is present in the LMC \citep{narloch2022}, searches performed with Z=0.010 BPASS models revealed analogous pathways and so the specific choice of metallicity does not impact our results.
The initial mass function is a Kroupa prescription \citep{kroupa01} with maximum mass 300 \msol. 
The binary fraction and initial parameters of the binary models (mass ratio, period) are taken from the observational survey of \cite{moe17}.
SN explosions cannot be modelled in detail but BPASS data products include ejecta masses and remnant masses for three likely scenarios: a weak SN explosion (10$^{50}$ ergs), a typical SN (10$^{51}$ ergs), and a very energetic explosion (10$^{52}$ ergs). 
For the purposes of this investigation we performed numerical simulations and model searches for all three scenarios as BHs can originate from very powerful SNe (e.g. in the collapsar scenario \citealt{woosley1993, woosley2012}) or result in failed SNe \citep{smartt2009,reynolds2015}. 
These alternative scenarios will change the remnant and ejecta mass of the post-SN model which will only indirectly affect the final orbit when using reduced Hobbs or Ultra-stripped SN kick prescriptions (see below), but directly affect the kick value if using a \cite{bray2016} prescription where the ejecta mass and remnant mass are parameters included in the definition. 

In order to take into account the effect of SN kicks we use an auxiliary code called {\sc tui} (previously used and described, although not mentioned by name at the time, in \citealt{ghodla2022}).
In this work we implement new kick prescriptions which focus on the specific case of BH remnants. 
A common kick prescription is a reduced kick (e.g. \citealt{compas2021}) where the BH remnants are assumed to receive the same momentum as the neutron star remnants. 
In this scenario the kick value is scaled by 1.4/$M_{\rm BH}$ where 1.4 is taken as the typical mass of a neutron star. 
One of the most commonly used neutron star kick distribution is a Maxwellian with $\sigma=265$\kms also called a Hobbs distribution \citep{hobbs2005}; additionally for low ejecta masses some compact object population synthesis models include the effects of Ultra-Stripped SNe \citep{tauris2013}, often with a default value of 30 \kms. 
We combine these considerations into a split kick prescription, where for a SN with an ejecta mass $>0.35$ \msol \citep{yao2020}, we use the BH mass scaled hobbs distribution, while below 0.35 \msol we impart a kick of 30\kms which we also scale by the mass of the remnant. 
In the final matches we find to VFTS 243, no progenitor matched the Ultra-stripped SN criteria. 
To further investigate the kick distribution of VFTS 243-like systems without any prior assumptions, we also perform our analysis with kicks drawn from a uniform distribution between 0 and 100\kms -- note that no mass scaling is applied in this case.

For each model undergoing a SN (i.e. obeys the contidition that its CO core mass  $>1.38$ \msol) we draw 2000 kicks and record the models that remain bound.
In addition we also consider a rotational velocity criteria as the O star in VFTS 243 was inferred to have a spin velocity at least 180\kms (\citealt{shenar2022}).
The original STARS models in BPASS do not take into account rotation in the evolution of the stellar interior (although it is considered in the orbital evolution). 
Using a disk-to-star angular momentum accretion efficiency of $\nu=0.5$ (i.e. 50 percent) any star accreting mass $\Delta M > M r_{g}^{2} \frac{180 / \rm{v}_{ \rm cr}}{\nu }$ is considered to be spun up to velocity $>$180\kms. Here $M, r_{g}, \rm{v}_{\rm cr}$ are the pre-accretion mass, radius of gyration and the critical velocity of the star (respectively) and it is assumed that the star is rotating as a solid body. This expression has been adopted from \cite{ghodla2023} who studied the spin up efficiency in the secondary component of a binary system due to accretion of mass from a Keplerian disk around the accretor. Since the projection of the VFTS 243 orbit is unknown, we allow all velocities larger than 180 \kms (up to critical) in our working.


\section{Results}
\label{sec:results}
\subsection{Posterior kick velocities}

\begin{figure}
	\includegraphics[width=8.5cm]{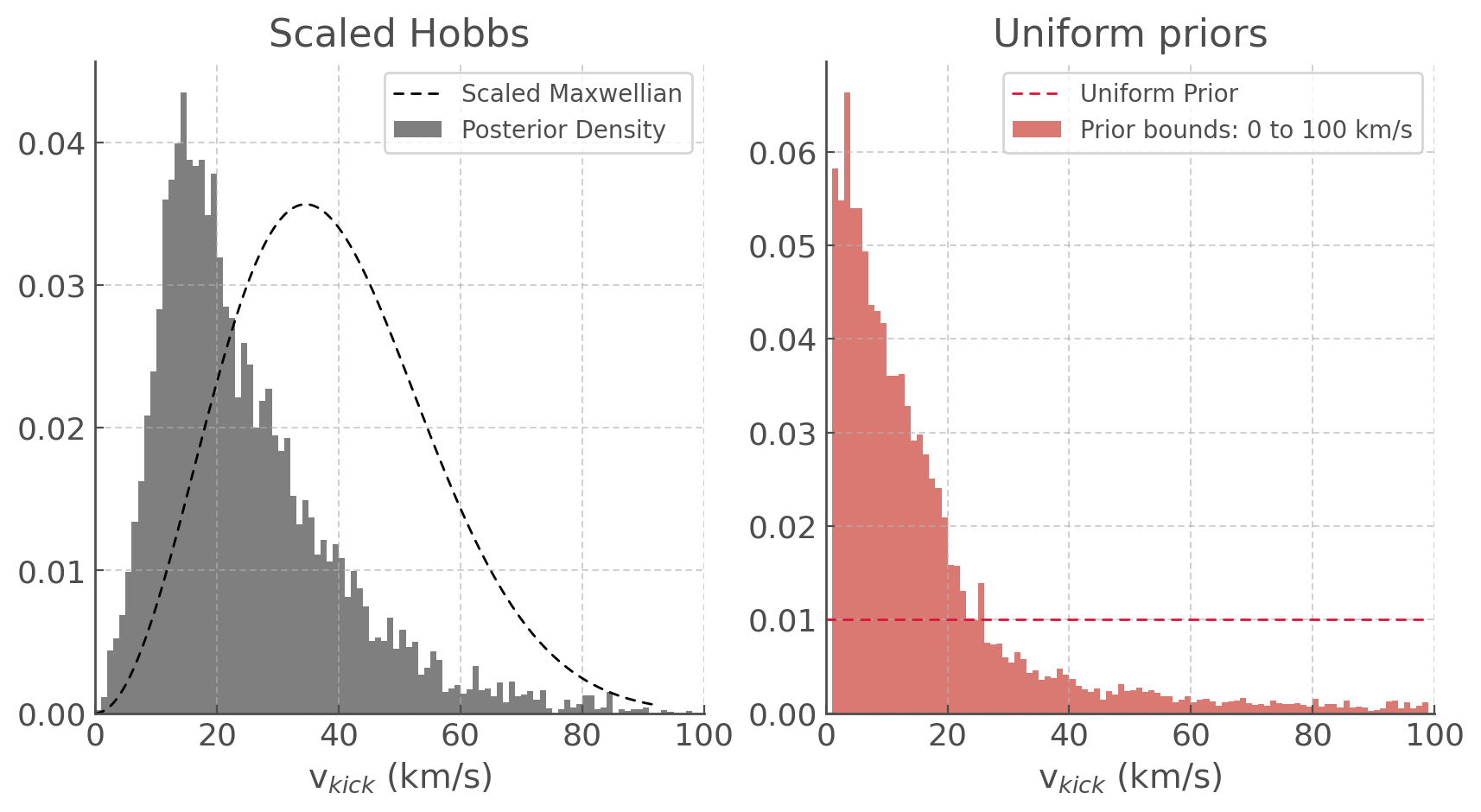}
    \caption{Posterior kick distributions of Star+BH systems with post SN periods between 9 and 11 days, and eccentricities lower than 0.05 for a BPASS fiducial population with metallicity Z=0.006. This is retrieved with 2000 random kick per SN and remnant masses calculated for weak SNe with E=10$^{50}$ ergs.}
    \label{fig:kick_psoteriors}
\end{figure}

In the next section we will search for the systems that are most similar to VFTS 243 but firstly, to explore the impact of kicks, we only consider the period and eccentricity criteria. 
\cite{shenar2022} find a period of 10.4 days and eccentricity of 0.017 $\pm\, 0.012$, additionally the rotation period they infer for the O star is not synchronised with the orbit and therefore tidal interactions (which would circularise the orbit) can be neglected. 
Consequently, we need to look for models which have similar orbital configuration after the first SN, without undergoing further binary interactions to re-process the orbit. 
Here we consider all models which have a period between 9 and 11 days and an eccentricity lower than 0.05, after SN kicks have been taken into account.

In Figure \ref{fig:kick_psoteriors} we show the kick velocities of the matched models from numerical simulations performed with a scaled Hobbs kick and the uniform kick distribution. 
We can see that the flat priors result in preferred kick velocities peaking at the lowest values (below 5\kms) and decreasing steadily -- 90 percent of the distribution is contained below 33\kms.
On the other hand the scaled Hobbs simulations peak between 10 and 20 \kms.
This is a result of the prior distribution which, with 2000 random kicks, still significantly influences the outcome given the very low weights of small kick velocities (even with the mass scaling).
To give a first order visual representation of the prior distribution in Figure \ref{fig:kick_psoteriors} \textit{after mass scaling}, we plot a Hobbs distribution scaled by the mean BH remnant mass in the sample on the x axis, and scaled by an arbitrary constant on the y axis to make it fit within the frame (the absolute height of the peak is meaningless but the shape is the crucial feature here). 
We can see that the posterior kick velocities peak \about 20\kms\ lower than the average prior (a factor of 2 lower), and the best kick velocities are from the uniform prior are a factor of \about 30 lower. 
Consequently we conclude that VFTS 243-like BH remnants do not receive the same momentum as the `typical' neutron star remnant (assuming the typical neutron star remnant kick is the Hobbs distribution -- see Section \ref{sec:discussion} for a discussion).
This is most probably due to the different nature of the result of the core-collapse and explosions yielding the different types of compact remnants.

\subsection{Genealogy of VFTS 243}

In addition to the orbital parameters we can further narrow down matching BPASS systems by using the mass constraints derived by \cite{shenar2022}.
Spectroscopic masses were inferred using CMFGEN (19.3 $\pm5.2$ \msol; \citealt{hillier2001}\footnote{\url{http://kookaburra.phyast.pitt.edu/hillier/web/CMFGEN.htm}}) as well as PoWR (22$\pm$5; \citealt{hamann2003, sander2015})\footnote{\url{https://www.astro.physik.uni-potsdam.de/~wrh/PoWR/powrgrid1.php}} and FASTWIND (13\msol; \citealt{puls2005,puls2020})\footnote{\url{https://fys.kuleuven.be/ster/research-projects/equation-folder/codes-folder/fastwind}}.
To be conservative with their interpretation they state a mass estimate of 25$\pm 12$\msol; later in the analysis when constraints on $M_*$ and $M_{\rm BH}$\footnote{We use M$_*$ to denote the mass of the bright O star and M$_{\rm BH}$ to denote the mass of the dark companion.}  from the orbital configuration are presented (see their Supplementary Figure 10), the masses for the O star are plotted as far as \about 27\msol. 
We therefore use a mass window between 13 and 27\msol\, for the O star.
We also include a total mass constraint of $<40$\msol\, to coincide with the total mass estimate of 36$^{+3.8}_{-5.4}$, as well as a mass ratio (q=$M_*$/$M_{\rm BH}$) between 1.5 and 3 (note that because of the mass ratio constrain a low mass limit on the total mass is redundant).
Finally we add a constraint on the radius of the O star to be within roughly 3\rsol of the value derived by \cite{shenar2022} of \about 10\rsol. 
Our search criteria are summarised below and the properties of the two best models are shown in Table~\ref{tab:best_models}:
\begin{enumerate}
    \item \textit{e} (post SN) $<$ 0.05
    \item P (days, post SN)   $\in$ (9,11)
    \item $M_*$+$M_{\rm BH}$ < 40 \msol
    \item $M_*$  $\in$ (13,27) \msol
    \item q $\in$ (1.5-3) 
    \item log(R$_*$) $\in$ (0.85,1.1)
\end{enumerate}

\begin{table*}       
  \caption{Properties of the BPASS models that most closely resemble VFTS 243.}
 \begin{tabular*}{12cm}{l l l l l l l l l l l l l l}
  \hline
  Model& M$_{\rm 1, i}$ & M$_{\rm 2, i}$ & P$_i$&  M$_{\rm 1, f}$ & M$_{\rm BH}$ & M$_*$ &  P$_f$ & \textit{e}  & M$_{\rm ej}$ & N&\\
   & (\msol) & (\msol) & (d) &  (\msol) & (\msol) & (\msol) & (d) & -- & (\msol) & per 10$^6$\msol\\
  \hline
  141048 & 50 & 25 & 15.9  & 13.8 & 11.6 & 25.9 & 9.2  & 0.037 & 0.99 & 0.8 \\
  141238 & 40 & 24  & 15.9  & 11.6 & 9.5 & 24.8 & 9.0 & 0.048 & 1.08 & 0.9 \\
  \hline
   \label{tab:best_models} 
 \end{tabular*}
\end{table*}

The masses of the O stars in the best BPASS matches are \about 25--26 \msol which is the same as was found using BONNSAI \citep{schneider2014} by \cite{shenar2022}.
The ages however differ: the BONNSAI models predicted an age of 3.9 Myrs whereas our systems are between 4.6 and 5.4 Myrs when the primary explodes as a SN. 
This is likely a direct consequence of the fact that BONNSAI does not take into account the effects of binary interactions. 
On the other hand the exemplar evolution calculated by \cite{shenar2022} using \textsc{mesa} die after 7.1 Myrs.
This the result of the lower initial masses in their models (M$_1$ = 30.1\msol\ and M$_2$ = 21.9\msol) compared to ours (see Table \ref{tab:best_models}) leading to longer lifetimes. 

Additionally, the two best models presented here are matched immediately after the first SN, meaning that we only take into account the lifetime of the primary. 
We can look for secondary models which match our criteria, however we find that these matches have ages \about 9-10 Myrs, which is a little too old for the 30 Doradus region.
Our two best models presented here on the other hand are very consistent with the peak of star formation as derived by \cite{schneider2018}.

\begin{figure*}
	\includegraphics[width=14cm]{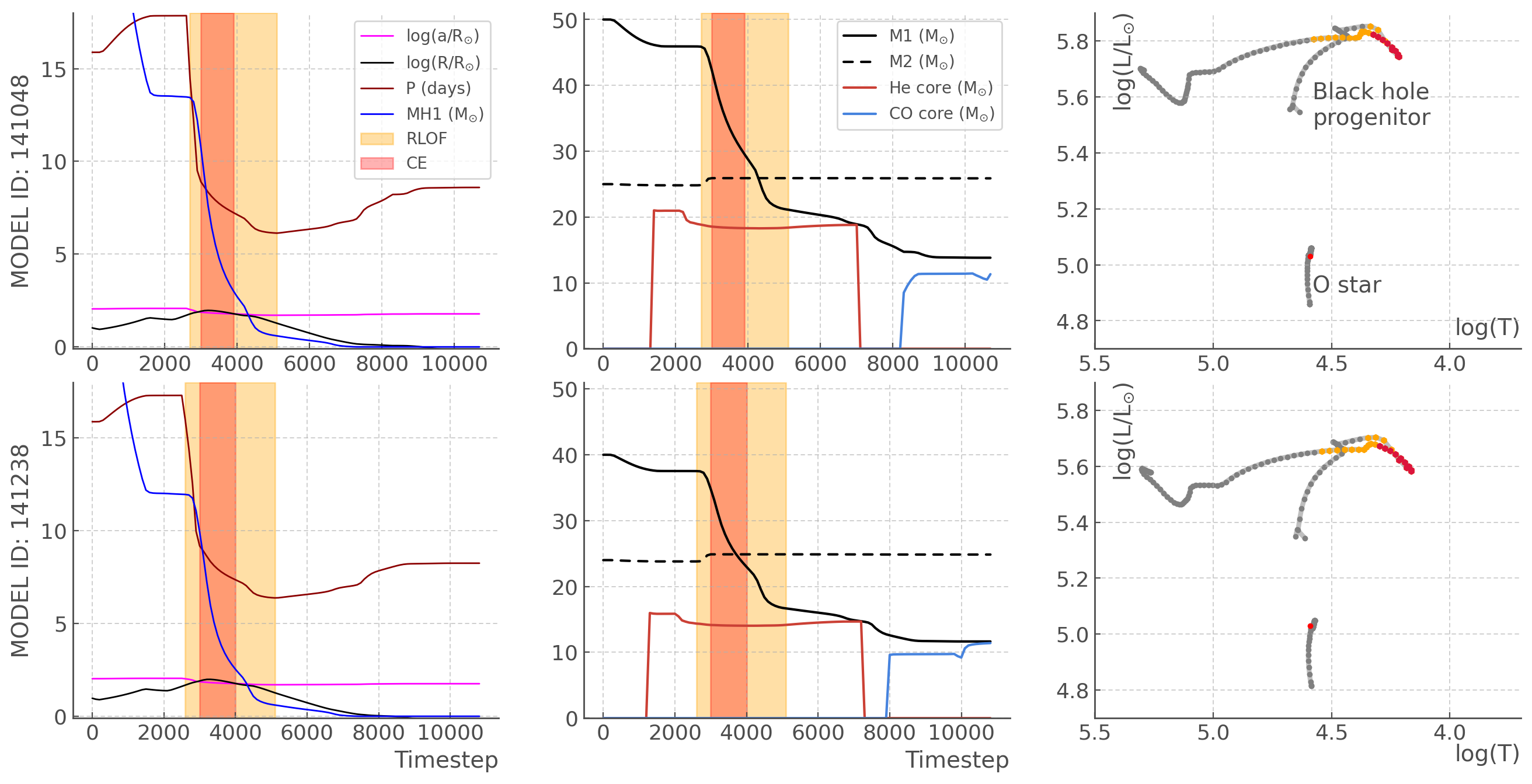}
    \caption{Evolution of the best progenitor candidates of VFTS 243 in BPASS at Z=0.006. Note that the time steps are not evenly spaced in time: the time intervals are dynamically chosen by the code to capture the changing behaviour of the star - in phases where the star changes a lot more steps are taken. This is a convenient way to visualise the changing parameters of a system as age alone can compress the information due to the short timescales of most critical phases of stellar evolution. Each point on the H-R Diagram is a time step in the model. Note that MH1 is the hydrogen envelope mass of the primary.}
    \label{fig:tracks}
\end{figure*}

The most striking difference between the exemplar model presented in \cite{shenar2022} and the ones we recover here is the type of mass transfer experienced by the systems. 
The original VFTS~243 study calls for a binary with an initially very low period (\about 3 days) which subsequently undergoes Case A (i.e. on the main sequence) stable mass transfer.
Our models on the other hand (see Figure \ref{fig:tracks}) go through a more typical Case B Roche Lobe Overflow. 
An episode of Common Envelope (CE)\footnote{Note that the CE criteria here is $R_*>a$} occurs after the onset of helium burning.
Both the separation and primary mass quickly decrease and the envelope collapses again below the CE threshold. 
The secondary gains very little mass in the process but the \about 20 solar masses lost by the primary in this interaction lead to a mass reversal. 
Further mass loss from the primary occurs when it becomes a helium star which leads to an increase in the period by a few days. 

Analogues for this system before the first SN exist within our Galaxy. 
The evolution we see here is similar to that found with earlier BPASS v1.0 models for the binary $\gamma^2$-Velorum \citep{eldridge2009}. 
That system has had a binary interaction in a slightly wider system with some mass transfer having occurred to the O star companion which shows indications of enhanced rotation. 
While VFTS 243 has a narrower orbit the highly non-conservative post-main sequence mass transfer event is observed for both systems

In the models presented here the death of the primary is a low energy (10$^{50}$ ergs) SN.
Often this is associated with Electron Capture SNe but in this mass range it can be thought of as the output of a partially failed SN. 
The BH remnant and fast rotation of the progenitor (given the binary interactions it underwent) could lead to the formation of a central engine whose jets, if it is not sufficiently long lived, may be choked by the envelope.
The choked jet model has been put forward as a unifying model for Long-Gamma-Ray-burst SNe and Ic-bl SNe by \cite{lazzati2012} (with some observational support from spectropolarimetry \citealt{stevance2017}), and it is conceivable that sufficient quenching would lead to lower kinetic energy release, although a quantitative approach is beyond the scope of this study.

In our model search we tried all three SN energy scenarios available in BPASS as it is unclear what type of energy we should expect \textit{a priori} from the collapse to a BH.
In this case the weak explosion scenario is the most successful at recreating VFTS~243; hypernova models are excluded completely and typical SNe are strongly disfavoured.
Furthermore, given that the ejecta mass for the 10$^{50}$ ergs explosions is \about 1 \msol, whereas the ejecta mass constraints in \cite{shenar2022} are $<0.5$ \msol, we can conclude that our models would in fact need to undergo an explosion with E$_{\rm KE}$ < 10$^{50}$ ergs. 
The fall back of ejected material onto the remnant would increase the remnant masses by a few tenths of \msol; our models would still be fully consistent with the observational criteria and fit a wider range of inclinations for VFTS~243 (see Supplementary Figure 10 in \citealt{shenar2022}).

\begin{figure}
	\includegraphics[width=8cm]{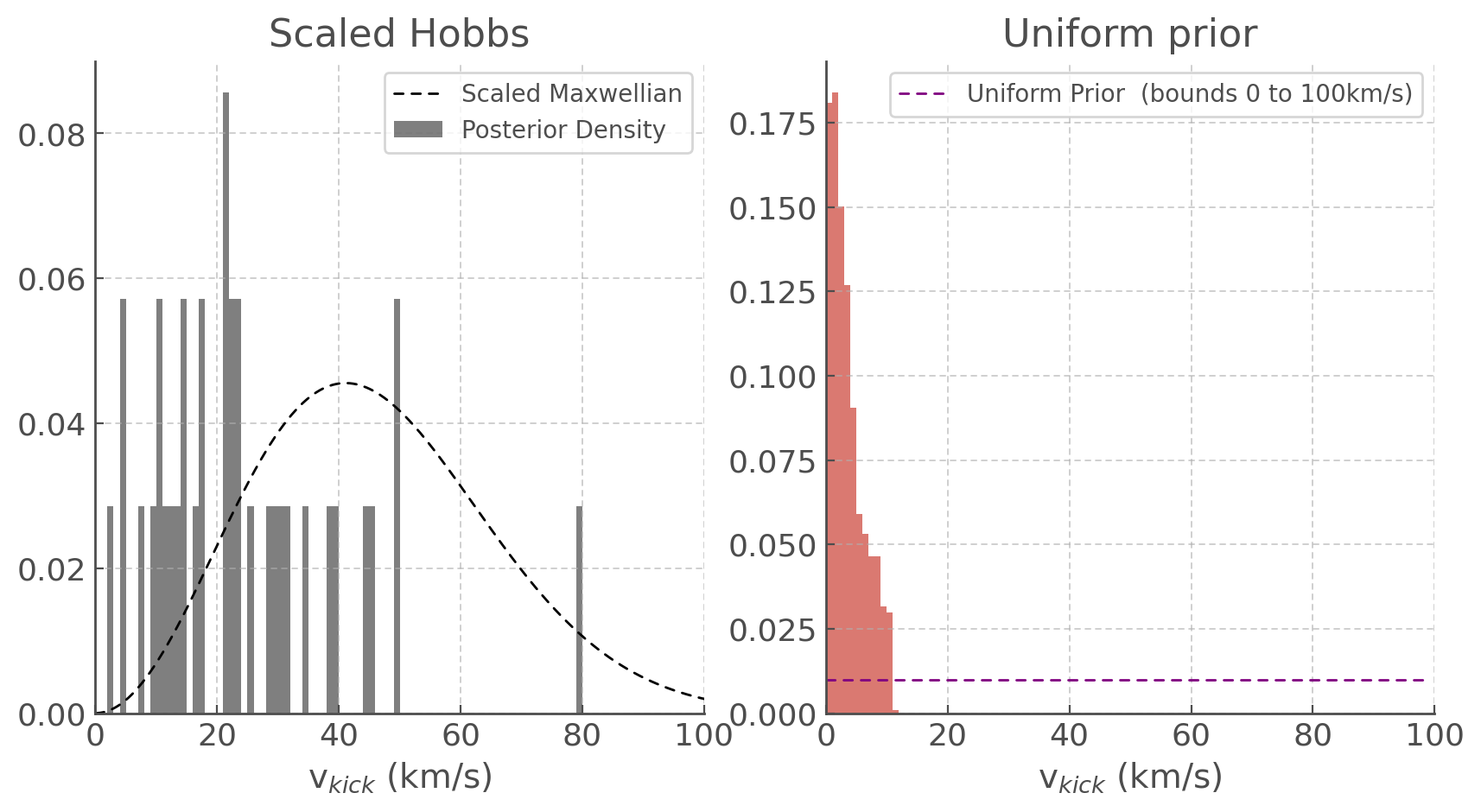}
    \caption{Sampled kick velocities for the two best models (including physical as well as orbital matching criteria). }
    \label{fig:vkick_best_mods}
\end{figure}

Finally we show the sample of kick velocities for the two best models highlighted in this section (see Figure \ref{fig:vkick_best_mods}). As we can see the Uniform prior samples show a very similar behaviour to that presented in Figure \ref{fig:kick_psoteriors} and very low velocities are preferred. 
The lack of samples beyond v$_{\rm kick}$ \about 10 \kms does not necessarily mean that these velocities are forbidden and could be the result of low number statistics, as here we have N=1073 matching models compared to over 8000 in Figure \ref{fig:kick_psoteriors}.
There are very few matching samples with a Hobbs kick distribution (N=36) and overall they are outweighed by the Uniform kick models by a factor of 30 (when taking into account he incidence rates of each system due to the initial mass function). 
Overall, considering the physical (mass and radius) constraints on VFTS 243 strengthens the conclusion that this system underwent an extremely low kick that is at odds with the traditional kick velocities assumed for black-hole remnants.

\section{Discussion}
\label{sec:discussion}

\begin{figure*}
	\includegraphics[width=16cm]{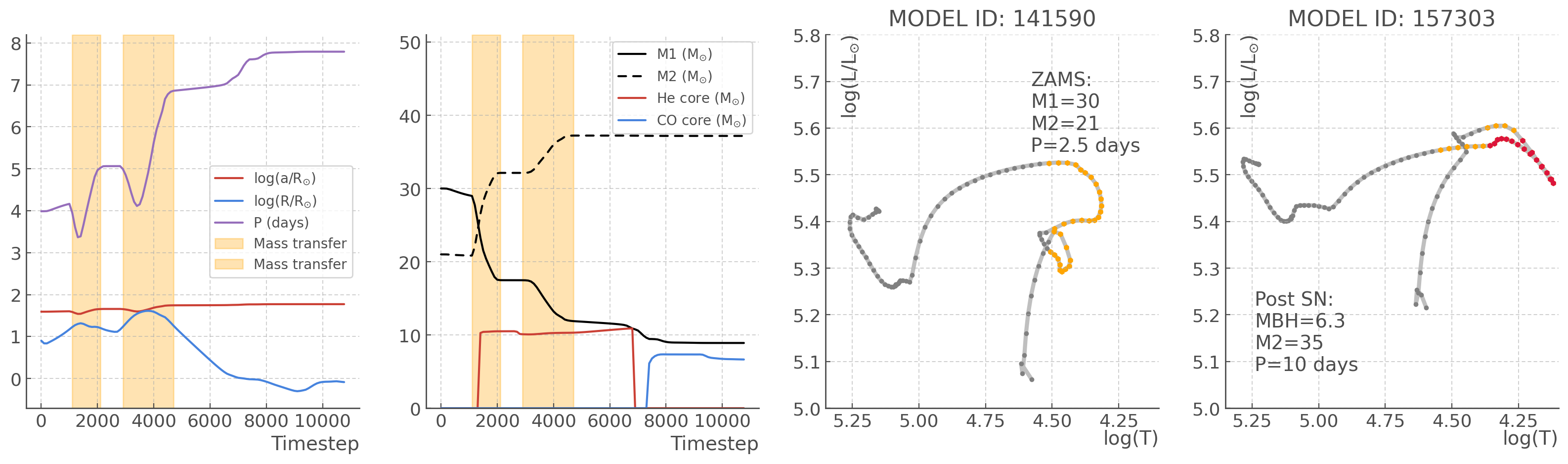}
    \caption{Closest BPASS genealogy to the examplar progenitor model of VFTS 243 presented by \protect\cite{shenar2022}. Each point on the H-R Diagram is a time step in the model. }
    \label{fig:caseAmodel}
\end{figure*}

To further investigate why the BPASS fiducial models do not naturally result in predictions similar to the system presented in \cite{shenar2022}, we specifically look for models with similar initial parameters. 
The discovery paper finds their best model to have $M_1=30.1$\msol, $M_2=21.9$\msol, and a period of 3.7 days.
On the BPASS model grid the nearest system is Model 141590, with initial $M_1=30$\msol, $M_2=21$\msol\,and a period of 2.5 days.
Its evolution is shown in Figure \ref{fig:caseAmodel}.

Like the \textsc{mesa} model, the BPASS model also undergoes Roche Lobe Over-flow (RLOF) on the main sequence (Case A), which leads to a significant exchange of mass between the primary and the companion, leading the secondary to reach \about 32\msol, which is already much greater than the mass estimates from the observations shown in the discovery paper. 
The primary briefly detaches after the onset of helium burning, then RLOF resumes and a further \about 5\msol\ are accreted. 
The end product is a 6.7\msol\ BH around a 37\msol\ star.
At this stage one might argue that future binary interaction could lead to sufficient mass loss in the secondary star (and mass increase of the BH) to result in a mass regime consistent with VFTS 243, however \cite{shenar2022} demonstrated that the O star in the system is not synchronised with the orbit and therefore no significant interaction has taken place between the companion and the BH. 

Overall, the BPASS equivalent to the examplar model in \cite{shenar2022} is not a good progenitor of VFTS 243. 
Stellar evolution models are complex and there may be several factors at play, but the most likely source of this difference between the BPASS and the \textsc{mesa} models is the mass transfer efficiency. 
The question of how much mass the secondary does accrete during a RLOF episode is not settled and in the \textsc{mesa} models the mass-transfer efficiency was treated as a free parameter. 
Their examplar model requires 64 percent of the mass transferred from the primary to be accreted by the secondary.
In BPASS, the detailed stellar models are calculated using the Cambridge STARS code where the mass transfer efficiency is not a free parameter and in the case of Case A mass transfer, which occurs on the nuclear time scale, the mass transfer efficiency\footnote{Calculated \textit{a posterori}} is very high (\about 97 percent in this particular case). 
This explains why the BPASS model with very similar initial parameters result in a system where the O star secondary is too massive to match VFTS 243. 

In terms of rate, we find that the models presented in Section \ref{sec:results} are expected to occur 0.8 and 0.9 times (respectively) in a 1 million \msol\, population. 
The slightly lower primary mass in the {\sc mesa} analogue presented here leads to a slightly higher rate: \about 1.2 per millions \msol. 
However the star formation history of 30 Doradus indicates a  higher star formation rate around 5 Myrs ago than 7 Myrs ago  by roughly a factor of \about 1.5 (see \citealt{schneider2018}). 
Overall, occurance rates alone do not provide clear support to one or the other pathway. 
A better understanding of stable mass transfer efficiency (particularly on the main sequence) will be required to elucidate which channel is better suited.

Another very uncertain phase in stellar evolution model is the CE.
The treatment of CE in BPASS differs from the classical approach (Stevance et al. in prep); one important point is that the CE prescription in our STARS implementation does not systematically require the entire envelope to be expelled.
This can be seen in the left-most panels of Figure \ref{fig:tracks}: $>0.5$\msol\, of hydrogen remains in the envelope when the system detaches. 
Additionally we conserve angular momentum rather than binding energy which overall makes our CE more efficient at unbinding the envelope, leading to less orbital shrinkage. 
A different CE prescription from the one we use would likely lead to very different final systems when using the same initial parameters as those reported in Table \ref{tab:best_models}, and a mismatch with VFTS~243. 

The fiducial BPASS populations have been able to reproduce a number of observable phenomena particularly as they relate to massive stars.
Nevertheless, CE remains a very uncertain phase of evolution and this should be kept in mind whenever specific evolutionary channels are being studied, as opposed to large populations.
In particular we note here the CE phase for our favoured models matching VFTS~243 occurs very shortly after the main sequence of the primary. 
Such Case B interactions in the Hertzsprung gap are uncertain; in rapid population synthesis models the implementation used is varied as it can have impacts on predictions (e.g. \citealt{belczynski2016}).
In BPASS as we can see such a CE leads to significant mass loss and period decrease but with the orbital radius being mainly unaffected. 
Most of the orbital evolution occurs before the CE due to mass transfer. This evolution is relatively robust against uncertainties in the CE model, because if CE was to have weaker/stronger orbital evolution than we predict here there are always narrower/wider orbits to provide the same eventual endpoint respectively. 

Another very uncertain aspect of the physics of compact object creation is the natal kicks they undergo. 
Neutron star kicks have empirical values based on runaway pulsar velocities (e.g. \citealt{hobbs2005,verbunt2017}) which are often used in compact object population synthesis. 
The velocity distribution of a bound neutron star is expected to be lower however, and there is evidence to support this (e.g. \citealt{tauris2017,abbott2017a,vigna-gomez2018}, Stevance et al. in prep). 
As for BH natal kicks, past studies of X-ray binaries have demonstrated that their natal kicks are lower than those of neutron star: between 80 and 310 \kms for XTE J1118+480 \citep{fargos2009}; $<80 $\kms for Cygnus X-1 \citep{nelemans1999}; and of order a few tens of \kms for GRO J1655-40 \citep{willems2005}.
Follow-up analysis by \cite{mandel2016} demonstrated that these observations did not require natal kicks much greater than 80\kms\ (although higher kicks could not be ruled out). 
Due to the lack of BH specific prescription it is common in modelling to adopt a reduced version of the neutron star kicks (see Section \ref{sec:methods}).

In this study we showed that the posterior kick velocity of our matching BPASS models peaks at values significantly lower than one would expect from this reduced kick. 
From numerical simulations using uniform priors we find that the BH in VFTS 243 likely received a kick $<$33\kms with 90 percent confidence.
This is consistent with the finding that our models require a low explosion energy ($<10^{50}$ ergs).
It is worth considering what change in velocity we might expect the BH to undergo purely from the instantaneous mass-loss.
The orbital velocity of the less massive star in a binary system is (following the notation of \citealt{blaauw1961}): 
\begin{equation}
S_2 = 30\times\frac{M1}{M1+M2}\times\sqrt{\frac{M_2}{a} + \frac{M_1}{a}}
\end{equation}
where $M_1$ and $M_2$ are the masses of the most and least massive stellar components respectively, and $a$ is the separation between the two stars in astronomical units. 
For our best models, we find that the orbital velocity changes by $<2$ \kms\ after the death of the primary.
This is consistent with the peak of the posterior velocity distribution obtained from our simulations performed with the uniform prior (see Figure \ref{fig:kick_psoteriors}), although the tail of this distribution extends past values consistent with this `pure Blaauw' kick.
Given the very low eccentricity of VFTS~243 however, it is very likely that the asymmetrical kick velocity was indeed very small.

It is important to note that the Maxwellian kicks and Ultra-stripped SNe kicks mentioned in this work are not the only prescriptions in the literature. 
For example \cite{baker2008} presented numerical simulations investigating the perpendicular kick as a function of $\eta =(q/(1+q))^2$, and \cite{mandel2020} developed analytical prescriptions based on the CO core and helium shell masses (which also stand out by their probabilistic nature as opposed to being deterministic like most prescriptions).
In another approach presented by \cite{bray2016} the natal kick values are a function of the mass ejected in the SN explosion. 
Most recently Richards et al. (2022, submit.) refined the $\alpha$, $\beta$ parameters involved by comparing to a larger range of observables.  
In this prescription, the kick velocity is given by

\begin{equation}
    v_\text{k} = \alpha\left(\frac{m_\text{ej}}{m_\text{rem}}\right)+\beta\left(\frac{1.4\msol}{m_\text{rem}}\right),
\end{equation}

where $\alpha$ and $\beta$ are constants.
We can use their quoted best parameters ($\alpha=115^{+35}_{-50}\kms$; $\beta = 15^{+10}_{-15}\kms$), the ejecta mass ($<0.5$\msol) and the remnant mass for VFTS~243 ($10.1\pm2$\msol) to calculate the reduced Bray kick (as opposed the a reduced Hobbs kick, see Section \ref{sec:methods}).
We find a kick velocity of 15.8 $\pm 2.7$ km s$^{-1}$, which is a slight overprediction of the peak of the posterior kick distribution obtained from the uniform prior in Section \ref{sec:results} (see Figure \ref{fig:kick_psoteriors}).
We note that that the best parameters obtained by Richards et al. (2023, in press.) are calibrated against the Galactic double neutron-star system population compiled in \cite{vigna-gomez2018} and a catalogue of single-star pulsars from \cite{willcox2021}; the comparison between this prescription and VFTS 243 is encouraging as to its potential to apply the BH population as well, but future research will be required to firmly establish its wider applicability.
Future discoveries and studies of quiescent BH binaries will therefore be essential for the further refinement of BH natal kicks, which is crucial as they play in important role in the fate of these systems \citep{dominik2012}.

\section{Summary and Conclusions}
\label{sec:conclusion}
In this work we conducted a search through the BPASSv2.2.1 of the recently reported VFTS~243 system containing an O7 star around a quiescent BH in a 10 day orbit at low eccentricity.
We investigated three prior kick distributions including the classic Hobbs distributions (Maxwellian with $\sigma=265$ \kms) and a low-kick uniform distributions ( 0--100 \kms), as well as three SN explosion energies (10$^{50}$, 10$^{51}$, 10$^{52}$ ergs). 
Our key findings are as follows:
\begin{enumerate}
    \item In BPASS, for a metallicity of $Z=0.006$, the most likely progenitors systems have primary mass 40-50\msol\ with secondaries \about 24-25\msol\ in \about 15 day orbits. 
    
    \item Roche Lobe Overflow occurs after the end of the main sequence and the mass transfer is highly non-conservative (accretion efficiency \about 5 percent). 
    
    \item The death of the BH progenitor resulted in an explosion with E$<10^{50}$ergs.
    This could be a good candidate for a failed SN, or, given the evidence for binary interaction and rapid rotation in the O star, a very short lived collapsar leading to choked jets and a low energy explosion and ejecta mass (\about 0.5\msol\ \citealt{shenar2022}). 
    
    \item The numerical simulations performed with the uniform prior revealed that black hole-O star systems with similar orbital parameters as VFTS 243 exhibited low SN kicks ($<33$\kms in the 90 percent credible interval) with the peak of the posterior distribution consistent with the magnitude of the Blaauw kick (solely the result of instantaneous mass loss). This low kick velocity requirement is even stronger in the models that best match VFTS 243, with a strong peak at 1-2\kms and v$\le$10\kms. The very low eccentricity reported for VFTS~243 and very low explosion energy required are also consistent with a scenario with a very low SN kick.
    
    \item The Hobbs kick priors are strongly disfavoured which is consistent with the low SN energy evidence. Indeed the Hobbs kicks are based on the velocities of runaway pulsars, which are very unlikely to be the result of low energy explosions.
    
    \item The theoretical kick prescriptions based on ejecta masses (\citealt{bray2016}, Richards et al. (2023, in press) predicts a kick velocity of 15.8 $\pm$ 2.7 \kms, which is only a slight overprediction of the results obtained from uniform priors.
\end{enumerate}

Further study of  quiescent BHs will allow further refinement of the BH natal kick distribution and a better understanding of mass transfer in their progenitor systems.
Other candidates found in VFTS \citep{shenar2022b} and in Gaia DR3 \citep{gaia2022} will be the object of a future study, and more observations of such systems are crucial to validating and constraining stellar evolution models at a key milestone in their journey towards becoming BH-BH merger progenitor systems.

\section*{Acknowledgements}
HFS and JJE acknowledge the support of the Marsden Fund Council managed through Royal Society Te Ap\=arangi. SMR, SG and PT acknowledge support from The University of Auckland. MMB acknowledges the support of the RSNZ.

\section*{Data Availability}
The BPASSv2.2.1 (and above) data products are available freely -- see \url{https://bpass.auckland.ac.nz/}. 
The {\sc tui} numerical simulations for this work and the codes to perform the analysis and create the plots can be requested by email to the first author.
The BPASS python toolkit hoki has been made Open Source \citep{stevance2020}. See \url{https://heloises.github.io/hoki/quick_start.html} for download instructions.



\bibliographystyle{mnras}

\begin{thebibliography}{}
\makeatletter
\relax
\def\mn@urlcharsother{\let\do\@makeother \do\$\do\&\do\#\do\^\do\_\do\%\do\~}
\def\mn@doi{\begingroup\mn@urlcharsother \@ifnextchar [ {\mn@doi@}
  {\mn@doi@[]}}
\def\mn@doi@[#1]#2{\def\@tempa{#1}\ifx\@tempa\@empty \href
  {http://dx.doi.org/#2} {doi:#2}\else \href {http://dx.doi.org/#2} {#1}\fi
  \endgroup}
\def\mn@eprint#1#2{\mn@eprint@#1:#2::\@nil}
\def\mn@eprint@arXiv#1{\href {http://arxiv.org/abs/#1} {{\tt arXiv:#1}}}
\def\mn@eprint@dblp#1{\href {http://dblp.uni-trier.de/rec/bibtex/#1.xml}
  {dblp:#1}}
\def\mn@eprint@#1:#2:#3:#4\@nil{\def\@tempa {#1}\def\@tempb {#2}\def\@tempc
  {#3}\ifx \@tempc \@empty \let \@tempc \@tempb \let \@tempb \@tempa \fi \ifx
  \@tempb \@empty \def\@tempb {arXiv}\fi \@ifundefined
  {mn@eprint@\@tempb}{\@tempb:\@tempc}{\expandafter \expandafter \csname
  mn@eprint@\@tempb\endcsname \expandafter{\@tempc}}}

\bibitem[\protect\citeauthoryear{Abbott et~al.,}{Abbott
  et~al.}{2017}]{abbott2017a}
Abbott B.~P.,  et~al., 2017, \mn@doi [ApJL] {10.3847/2041-8213/aa93fc}, 850,
  L40

\bibitem[\protect\citeauthoryear{{Baker}, {Boggs}, {Centrella}, {Kelly},
  {McWilliams}, {Miller}  \& {van Meter}}{{Baker} et~al.}{2008}]{baker2008}
{Baker} J.~G.,  {Boggs} W.~D.,  {Centrella} J.,  {Kelly} B.~J.,  {McWilliams}
  S.~T.,  {Miller} M.~C.,   {van Meter} J.~R.,  2008, \mn@doi [\apjl]
  {10.1086/590927}, \href
  {https://ui.adsabs.harvard.edu/abs/2008ApJ...682L..29B} {682, L29}

\bibitem[\protect\citeauthoryear{Belczynski, Holz, Bulik  \&
  O'Shaughnessy}{Belczynski et~al.}{2016}]{belczynski2016}
Belczynski K.,  Holz D.~E.,  Bulik T.,   O'Shaughnessy R.,  2016, \mn@doi
  [Nature] {10.1038/nature18322}, 534, 512

\bibitem[\protect\citeauthoryear{Blaauw}{Blaauw}{1961}]{blaauw1961}
Blaauw A.,  1961, Bulletin of the Astronomical Institutes of the Netherlands,
  15, 265

\bibitem[\protect\citeauthoryear{{Bodensteiner} et~al.,}{{Bodensteiner}
  et~al.}{2020}]{bodensteiner2020}
{Bodensteiner} J.,  et~al., 2020, \mn@doi [\aap] {10.1051/0004-6361/202038682},
  \href {https://ui.adsabs.harvard.edu/abs/2020A&A...641A..43B} {641, A43}

\bibitem[\protect\citeauthoryear{{Bodensteiner} et~al.,}{{Bodensteiner}
  et~al.}{2022}]{bodensteiner2022}
{Bodensteiner} J.,  et~al., 2022, arXiv e-prints, \href
  {https://ui.adsabs.harvard.edu/abs/2022arXiv220700366B} {p. arXiv:2207.00366}

\bibitem[\protect\citeauthoryear{{Bray} \& {Eldridge}}{{Bray} \&
  {Eldridge}}{2016}]{bray2016}
{Bray} J.~C.,  {Eldridge} J.~J.,  2016, \mn@doi [\mnras]
  {10.1093/mnras/stw1275}, \href
  {https://ui.adsabs.harvard.edu/abs/2016MNRAS.461.3747B} {461, 3747}

\bibitem[\protect\citeauthoryear{Briel, Stevance  \& Eldridge}{Briel
  et~al.}{2022}]{briel2022}
Briel M.~M.,  Stevance H.~F.,   Eldridge J.~J.,  2022, Understanding the
  High-Mass Binary Black Hole Population from Stable Mass Transfer and
  Super-{{Eddington}} Accretion in {{BPASS}}

\bibitem[\protect\citeauthoryear{{Byrne}, {Stanway}  \& {Eldridge}}{{Byrne}
  et~al.}{2021}]{byrne2021}
{Byrne} C.~M.,  {Stanway} E.~R.,   {Eldridge} J.~J.,  2021, \mn@doi [\mnras]
  {10.1093/mnras/stab2115}, \href
  {https://ui.adsabs.harvard.edu/abs/2021MNRAS.507..621B} {507, 621}

\bibitem[\protect\citeauthoryear{COMPAS et~al.,}{COMPAS
  et~al.}{2021}]{compas2021}
COMPAS T.,  et~al., 2021, arXiv:2109.10352 [astro-ph]

\bibitem[\protect\citeauthoryear{{Chrimes} et~al.,}{{Chrimes}
  et~al.}{2022}]{chrimes2022}
{Chrimes} A.~A.,  et~al., 2022, \mn@doi [\mnras] {10.1093/mnras/stac1796},
  \href {https://ui.adsabs.harvard.edu/abs/2022MNRAS.tmp.1741C} {}

\bibitem[\protect\citeauthoryear{Dominik, Belczynski, Fryer, Holz, Berti,
  Bulik, Mandel  \& O'Shaughnessy}{Dominik et~al.}{2012}]{dominik2012}
Dominik M.,  Belczynski K.,  Fryer C.,  Holz D.~E.,  Berti E.,  Bulik T.,
  Mandel I.,   O'Shaughnessy R.,  2012, \mn@doi [The Astrophysical Journal]
  {10.1088/0004-637X/759/1/52}, 759, 52

\bibitem[\protect\citeauthoryear{Eggleton}{Eggleton}{1971}]{eggleton1971}
Eggleton P.~P.,  1971, \mn@doi [Monthly Notices of the Royal Astronomical
  Society] {10.1093/mnras/151.3.351}, 151, 351

\bibitem[\protect\citeauthoryear{{El-Badry} \& Burdge}{{El-Badry} \&
  Burdge}{2021}]{el-badry2021}
{El-Badry} K.,  Burdge K.,  2021, arXiv:2111.07925 [astro-ph]

\bibitem[\protect\citeauthoryear{{El-Badry} \& {Quataert}}{{El-Badry} \&
  {Quataert}}{2020}]{el-badry2020}
{El-Badry} K.,  {Quataert} E.,  2020, \mn@doi [\mnras]
  {10.1093/mnrasl/slaa004}, \href
  {https://ui.adsabs.harvard.edu/abs/2020MNRAS.493L..22E} {493, L22}

\bibitem[\protect\citeauthoryear{{Eldridge}}{{Eldridge}}{2009}]{eldridge2009}
{Eldridge} J.~J.,  2009, \mn@doi [\mnras] {10.1111/j.1745-3933.2009.00753.x},
  \href {https://ui.adsabs.harvard.edu/abs/2009MNRAS.400L..20E} {400, L20}

\bibitem[\protect\citeauthoryear{{Eldridge}, {Izzard}  \& {Tout}}{{Eldridge}
  et~al.}{2008}]{eldridge08}
{Eldridge} J.~J.,  {Izzard} R.~G.,   {Tout} C.~A.,  2008, \mn@doi [\mnras]
  {10.1111/j.1365-2966.2007.12738.x}, \href
  {http://adsabs.harvard.edu/abs/2008MNRAS.384.1109E} {384, 1109}

\bibitem[\protect\citeauthoryear{Eldridge, Stanway, Xiao, McClelland, Taylor,
  Ng, Greis  \& Bray}{Eldridge et~al.}{2017}]{eldridge2017}
Eldridge J.~J.,  Stanway E.~R.,  Xiao L.,  McClelland L. A.~S.,  Taylor G.,  Ng
  M.,  Greis S. M.~L.,   Bray J.~C.,  2017, \mn@doi [Publications of the
  Astronomical Society of Australia] {10.1017/pasa.2017.51}, 34, e058

\bibitem[\protect\citeauthoryear{{Fragos}, {Willems}, {Kalogera}, {Ivanova},
  {Rockefeller}, {Fryer}  \& {Young}}{{Fragos} et~al.}{2009}]{fargos2009}
{Fragos} T.,  {Willems} B.,  {Kalogera} V.,  {Ivanova} N.,  {Rockefeller} G.,
  {Fryer} C.~L.,   {Young} P.~A.,  2009, \mn@doi [\apj]
  {10.1088/0004-637X/697/2/1057}, \href
  {https://ui.adsabs.harvard.edu/abs/2009ApJ...697.1057F} {697, 1057}

\bibitem[\protect\citeauthoryear{{Gaia Collaboration} et~al.,}{{Gaia
  Collaboration} et~al.}{2022}]{gaia2022}
{Gaia Collaboration} et~al., 2022, arXiv e-prints, \href
  {https://ui.adsabs.harvard.edu/abs/2022arXiv220605595G} {p. arXiv:2206.05595}

\bibitem[\protect\citeauthoryear{{Ghodla}, {van Zeist}, {Eldridge}, {Stevance}
  \& {Stanway}}{{Ghodla} et~al.}{2022}]{ghodla2022}
{Ghodla} S.,  {van Zeist} W. G.~J.,  {Eldridge} J.~J.,  {Stevance} H.~F.,
  {Stanway} E.~R.,  2022, \mn@doi [\mnras] {10.1093/mnras/stac120}, \href
  {https://ui.adsabs.harvard.edu/abs/2022MNRAS.511.1201G} {511, 1201}

\bibitem[\protect\citeauthoryear{{Ghodla}, {Eldridge}, {Stanway}  \&
  {Stevance}}{{Ghodla} et~al.}{2023}]{ghodla2023}
{Ghodla} S.,  {Eldridge} J.~J.,  {Stanway} E.~R.,   {Stevance} H.~F.,  2023,
  \mn@doi [\mnras] {10.1093/mnras/stac3177}, \href
  {https://ui.adsabs.harvard.edu/abs/2023MNRAS.518..860G} {518, 860}

\bibitem[\protect\citeauthoryear{{Hamann} \& {Gr{\"a}fener}}{{Hamann} \&
  {Gr{\"a}fener}}{2003}]{hamann2003}
{Hamann} W.~R.,  {Gr{\"a}fener} G.,  2003, \mn@doi [\aap]
  {10.1051/0004-6361:20031308}, \href
  {https://ui.adsabs.harvard.edu/abs/2003A&A...410..993H} {410, 993}

\bibitem[\protect\citeauthoryear{{Hillier} \& {Lanz}}{{Hillier} \&
  {Lanz}}{2001}]{hillier2001}
{Hillier} D.~J.,  {Lanz} T.,  2001, in {Ferland} G.,  {Savin} D.~W.,  eds,
  Astronomical Society of the Pacific Conference Series Vol. 247, Spectroscopic
  Challenges of Photoionized Plasmas. p.~343

\bibitem[\protect\citeauthoryear{Hobbs, Lorimer, Lyne  \& Kramer}{Hobbs
  et~al.}{2005}]{hobbs2005}
Hobbs G.,  Lorimer D.~R.,  Lyne A.~G.,   Kramer M.,  2005, \mn@doi [Monthly
  Notices of the Royal Astronomical Society]
  {10.1111/j.1365-2966.2005.09087.x}, 360, 974

\bibitem[\protect\citeauthoryear{{Kroupa}}{{Kroupa}}{2001}]{kroupa01}
{Kroupa} P.,  2001, \mn@doi [\mnras] {10.1046/j.1365-8711.2001.04022.x}, \href
  {https://ui.adsabs.harvard.edu/abs/2001MNRAS.322..231K} {322, 231}

\bibitem[\protect\citeauthoryear{{Lazzati}, {Morsony}, {Blackwell}  \&
  {Begelman}}{{Lazzati} et~al.}{2012}]{lazzati2012}
{Lazzati} D.,  {Morsony} B.~J.,  {Blackwell} C.~H.,   {Begelman} M.~C.,  2012,
  \mn@doi [\apj] {10.1088/0004-637X/750/1/68}, \href
  {https://ui.adsabs.harvard.edu/abs/2012ApJ...750...68L} {750, 68}

\bibitem[\protect\citeauthoryear{{Liu} et~al.,}{{Liu} et~al.}{2019}]{liu2019}
{Liu} J.,  et~al., 2019, \mn@doi [\nat] {10.1038/s41586-019-1766-2}, \href
  {https://ui.adsabs.harvard.edu/abs/2019Natur.575..618L} {575, 618}

\bibitem[\protect\citeauthoryear{{Mandel}}{{Mandel}}{2016}]{mandel2016}
{Mandel} I.,  2016, \mn@doi [\mnras] {10.1093/mnras/stv2733}, \href
  {https://ui.adsabs.harvard.edu/abs/2016MNRAS.456..578M} {456, 578}

\bibitem[\protect\citeauthoryear{Mandel \& Broekgaarden}{Mandel \&
  Broekgaarden}{2021}]{mandel2021}
Mandel I.,  Broekgaarden F.~S.,  2021, arXiv:2107.14239 [astro-ph]

\bibitem[\protect\citeauthoryear{Mandel \& M{\"u}ller}{Mandel \&
  M{\"u}ller}{2020}]{mandel2020}
Mandel I.,  M{\"u}ller B.,  2020, \mn@doi [Monthly Notices of the Royal
  Astronomical Society] {10.1093/mnras/staa3043}, 499, 3214

\bibitem[\protect\citeauthoryear{{Massey}, {Neugent}, {Dorn-Wallenstein},
  {Eldridge}, {Stanway}  \& {Levesque}}{{Massey} et~al.}{2021}]{massey2021}
{Massey} P.,  {Neugent} K.~F.,  {Dorn-Wallenstein} T.~Z.,  {Eldridge} J.~J.,
  {Stanway} E.~R.,   {Levesque} E.~M.,  2021, \mn@doi [\apj]
  {10.3847/1538-4357/ac15f5}, \href
  {https://ui.adsabs.harvard.edu/abs/2021ApJ...922..177M} {922, 177}

\bibitem[\protect\citeauthoryear{{Moe} \& {Di Stefano}}{{Moe} \& {Di
  Stefano}}{2017}]{moe17}
{Moe} M.,  {Di Stefano} R.,  2017, \mn@doi [\apjs] {10.3847/1538-4365/aa6fb6},
  \href {https://ui.adsabs.harvard.edu/abs/2017ApJS..230...15M} {230, 15}

\bibitem[\protect\citeauthoryear{{Narloch} et~al.,}{{Narloch}
  et~al.}{2022}]{narloch2022}
{Narloch} W.,  et~al., 2022, arXiv e-prints, \href
  {https://ui.adsabs.harvard.edu/abs/2022arXiv220713153N} {p. arXiv:2207.13153}

\bibitem[\protect\citeauthoryear{{Nelemans}, {Tauris}  \& {van den
  Heuvel}}{{Nelemans} et~al.}{1999}]{nelemans1999}
{Nelemans} G.,  {Tauris} T.~M.,   {van den Heuvel} E.~P.~J.,  1999, \aap, \href
  {https://ui.adsabs.harvard.edu/abs/1999A&A...352L..87N} {352, L87}

\bibitem[\protect\citeauthoryear{{Paxton}, {Bildsten}, {Dotter}, {Herwig},
  {Lesaffre}  \& {Timmes}}{{Paxton} et~al.}{2011}]{Paxton2011}
{Paxton} B.,  {Bildsten} L.,  {Dotter} A.,  {Herwig} F.,  {Lesaffre} P.,
  {Timmes} F.,  2011, \mn@doi [\apjs] {10.1088/0067-0049/192/1/3}, \href
  {https://ui.adsabs.harvard.edu/abs/2011ApJS..192....3P} {192, 3}

\bibitem[\protect\citeauthoryear{{Paxton} et~al.,}{{Paxton}
  et~al.}{2013}]{Paxton2013}
{Paxton} B.,  et~al., 2013, \mn@doi [\apjs] {10.1088/0067-0049/208/1/4}, \href
  {https://ui.adsabs.harvard.edu/abs/2013ApJS..208....4P} {208, 4}

\bibitem[\protect\citeauthoryear{{Paxton} et~al.,}{{Paxton}
  et~al.}{2015}]{Paxton2015}
{Paxton} B.,  et~al., 2015, \mn@doi [\apjs] {10.1088/0067-0049/220/1/15}, \href
  {https://ui.adsabs.harvard.edu/abs/2015ApJS..220...15P} {220, 15}

\bibitem[\protect\citeauthoryear{{Paxton} et~al.,}{{Paxton}
  et~al.}{2018}]{Paxton2018}
{Paxton} B.,  et~al., 2018, \mn@doi [\apjs] {10.3847/1538-4365/aaa5a8}, \href
  {https://ui.adsabs.harvard.edu/abs/2018ApJS..234...34P} {234, 34}

\bibitem[\protect\citeauthoryear{{Paxton} et~al.,}{{Paxton}
  et~al.}{2019}]{Paxton2019}
{Paxton} B.,  et~al., 2019, \mn@doi [\apjs] {10.3847/1538-4365/ab2241}, \href
  {https://ui.adsabs.harvard.edu/abs/2019ApJS..243...10P} {243, 10}

\bibitem[\protect\citeauthoryear{{Puls}, {Urbaneja}, {Venero}, {Repolust},
  {Springmann}, {Jokuthy}  \& {Mokiem}}{{Puls} et~al.}{2005}]{puls2005}
{Puls} J.,  {Urbaneja} M.~A.,  {Venero} R.,  {Repolust} T.,  {Springmann} U.,
  {Jokuthy} A.,   {Mokiem} M.~R.,  2005, \mn@doi [\aap]
  {10.1051/0004-6361:20042365}, \href
  {https://ui.adsabs.harvard.edu/abs/2005A&A...435..669P} {435, 669}

\bibitem[\protect\citeauthoryear{{Puls}, {Najarro}, {Sundqvist}  \&
  {Sen}}{{Puls} et~al.}{2020}]{puls2020}
{Puls} J.,  {Najarro} F.,  {Sundqvist} J.~O.,   {Sen} K.,  2020, \mn@doi [\aap]
  {10.1051/0004-6361/202038464}, \href
  {https://ui.adsabs.harvard.edu/abs/2020A&A...642A.172P} {642, A172}

\bibitem[\protect\citeauthoryear{{Reynolds}, {Fraser}  \& {Gilmore}}{{Reynolds}
  et~al.}{2015}]{reynolds2015}
{Reynolds} T.~M.,  {Fraser} M.,   {Gilmore} G.,  2015, \mn@doi [\mnras]
  {10.1093/mnras/stv1809}, \href
  {https://ui.adsabs.harvard.edu/abs/2015MNRAS.453.2885R} {453, 2885}

\bibitem[\protect\citeauthoryear{{Rivinius}, {Baade}, {Hadrava}, {Heida}  \&
  {Klement}}{{Rivinius} et~al.}{2020}]{rivinius2020}
{Rivinius} T.,  {Baade} D.,  {Hadrava} P.,  {Heida} M.,   {Klement} R.,  2020,
  \mn@doi [\aap] {10.1051/0004-6361/202038020}, \href
  {https://ui.adsabs.harvard.edu/abs/2020A&A...637L...3R} {637, L3}

\bibitem[\protect\citeauthoryear{Sana et~al.,}{Sana et~al.}{2012}]{sana2012}
Sana H.,  et~al., 2012, \mn@doi [Science] {10.1126/science.1223344}, 337, 444

\bibitem[\protect\citeauthoryear{Sana et~al.,}{Sana et~al.}{2014}]{sana2014}
Sana H.,  et~al., 2014, \mn@doi [The Astrophysical Journal Supplement Series]
  {10.1088/0067-0049/215/1/15}, 215, 15

\bibitem[\protect\citeauthoryear{{Sander}, {Shenar}, {Hainich},
  {G{\'\i}menez-Garc{\'\i}a}, {Todt}  \& {Hamann}}{{Sander}
  et~al.}{2015}]{sander2015}
{Sander} A.,  {Shenar} T.,  {Hainich} R.,  {G{\'\i}menez-Garc{\'\i}a} A.,
  {Todt} H.,   {Hamann} W.~R.,  2015, \mn@doi [\aap]
  {10.1051/0004-6361/201425356}, \href
  {https://ui.adsabs.harvard.edu/abs/2015A&A...577A..13S} {577, A13}

\bibitem[\protect\citeauthoryear{Saracino et~al.,}{Saracino
  et~al.}{2021}]{saracino2021}
Saracino S.,  et~al., 2021, arXiv:2111.06506 [astro-ph]

\bibitem[\protect\citeauthoryear{{Schneider}, {Langer}, {de Koter}, {Brott},
  {Izzard}  \& {Lau}}{{Schneider} et~al.}{2014}]{schneider2014}
{Schneider} F.~R.~N.,  {Langer} N.,  {de Koter} A.,  {Brott} I.,  {Izzard}
  R.~G.,   {Lau} H.~H.~B.,  2014, \mn@doi [\aap] {10.1051/0004-6361/201424286},
  \href {https://ui.adsabs.harvard.edu/abs/2014A&A...570A..66S} {570, A66}

\bibitem[\protect\citeauthoryear{{Schneider} et~al.,}{{Schneider}
  et~al.}{2018}]{schneider2018}
{Schneider} F.~R.~N.,  et~al., 2018, \mn@doi [Science]
  {10.1126/science.aan0106}, \href
  {https://ui.adsabs.harvard.edu/abs/2018Sci...359...69S} {359, 69}

\bibitem[\protect\citeauthoryear{{Shakura} \& {Sunyaev}}{{Shakura} \&
  {Sunyaev}}{1973}]{Sakura_Sunyaev_1973}
{Shakura} N.~I.,  {Sunyaev} R.~A.,  1973, \aap, \href
  {https://ui.adsabs.harvard.edu/abs/1973A&A....24..337S} {24, 337}

\bibitem[\protect\citeauthoryear{{Shenar} et~al.,}{{Shenar}
  et~al.}{2020}]{shenar2020}
{Shenar} T.,  et~al., 2020, \mn@doi [\aap] {10.1051/0004-6361/202038275}, \href
  {https://ui.adsabs.harvard.edu/abs/2020A&A...639L...6S} {639, L6}

\bibitem[\protect\citeauthoryear{Shenar et~al.,}{Shenar
  et~al.}{2022a}]{shenar2022}
Shenar T.,  et~al., 2022a, \mn@doi [Nat Astron] {10.1038/s41550-022-01730-y}

\bibitem[\protect\citeauthoryear{{Shenar} et~al.,}{{Shenar}
  et~al.}{2022b}]{shenar2022b}
{Shenar} T.,  et~al., 2022b, arXiv e-prints, \href
  {https://ui.adsabs.harvard.edu/abs/2022arXiv220707674S} {p. arXiv:2207.07674}

\bibitem[\protect\citeauthoryear{{Smartt}}{{Smartt}}{2009}]{smartt2009}
{Smartt} S.~J.,  2009, \mn@doi [\araa] {10.1146/annurev-astro-082708-101737},
  \href {https://ui.adsabs.harvard.edu/abs/2009ARA&A..47...63S} {47, 63}

\bibitem[\protect\citeauthoryear{Stanway \& Eldridge}{Stanway \&
  Eldridge}{2018}]{stanway2018}
Stanway E.~R.,  Eldridge J.~J.,  2018, \mn@doi [Monthly Notices of the Royal
  Astronomical Society] {10.1093/mnras/sty1353}, 479, 75

\bibitem[\protect\citeauthoryear{{Stanway}, {Eldridge}  \& {Chrimes}}{{Stanway}
  et~al.}{2020}]{stanway2020}
{Stanway} E.~R.,  {Eldridge} J.~J.,   {Chrimes} A.~A.,  2020, \mn@doi [\mnras]
  {10.1093/mnras/staa2089}, \href
  {https://ui.adsabs.harvard.edu/abs/2020MNRAS.497.2201S} {497, 2201}

\bibitem[\protect\citeauthoryear{{Stevance} \& {Eldridge}}{{Stevance} \&
  {Eldridge}}{2021}]{stevance2021}
{Stevance} H.~F.,  {Eldridge} J.~J.,  2021, \mn@doi [\mnras]
  {10.1093/mnrasl/slab039}, \href
  {https://ui.adsabs.harvard.edu/abs/2021MNRAS.504L..51S} {504, L51}

\bibitem[\protect\citeauthoryear{{Stevance} et~al.,}{{Stevance}
  et~al.}{2017}]{stevance2017}
{Stevance} H.~F.,  et~al., 2017, \mn@doi [\mnras] {10.1093/mnras/stx970}, \href
  {https://ui.adsabs.harvard.edu/abs/2017MNRAS.469.1897S} {469, 1897}

\bibitem[\protect\citeauthoryear{Stevance, Eldridge  \& Stanway}{Stevance
  et~al.}{2020}]{stevance2020}
Stevance H.,  Eldridge J.,   Stanway E.,  2020, \mn@doi [The Journal of Open
  Source Software] {10.21105/joss.01987}, 5, 1987

\bibitem[\protect\citeauthoryear{{Stevance}, {Parsons}  \&
  {Eldridge}}{{Stevance} et~al.}{2022}]{stevance2022}
{Stevance} H.~F.,  {Parsons} S.~G.,   {Eldridge} J.~J.,  2022, \mn@doi [\mnras]
  {10.1093/mnrasl/slac001}, \href
  {https://ui.adsabs.harvard.edu/abs/2022MNRAS.511L..77S} {511, L77}

\bibitem[\protect\citeauthoryear{Tang, Eldridge, Stanway  \& Bray}{Tang
  et~al.}{2020}]{tang2020}
Tang P.~N.,  Eldridge J.~J.,  Stanway E.~R.,   Bray J.~C.,  2020, \mn@doi
  [Monthly Notices of the Royal Astronomical Society] {10.1093/mnrasl/slz183},
  493, L6

\bibitem[\protect\citeauthoryear{Tauris, Langer, Moriya, Podsiadlowski, Yoon
  \& Blinnikov}{Tauris et~al.}{2013}]{tauris2013}
Tauris T.~M.,  Langer N.,  Moriya T.~J.,  Podsiadlowski P.,  Yoon S.-C.,
  Blinnikov S.~I.,  2013, \mn@doi [ApJ] {10.1088/2041-8205/778/2/L23}, 778, L23

\bibitem[\protect\citeauthoryear{Tauris et~al.,}{Tauris
  et~al.}{2017}]{tauris2017}
Tauris T.~M.,  et~al., 2017, \mn@doi [The Astrophysical Journal]
  {10.3847/1538-4357/aa7e89}, 846, 170

\bibitem[\protect\citeauthoryear{{The LVK Collaboration}}{{The LVK
  Collaboration}}{2021}]{theligoscientificcollaboration2021c}
{The LVK Collaboration} 2021, arXiv:2111.03634 [astro-ph, physics:gr-qc]

\bibitem[\protect\citeauthoryear{Verbunt, Igoshev  \& Cator}{Verbunt
  et~al.}{2017}]{verbunt2017}
Verbunt F.,  Igoshev A.,   Cator E.,  2017, \mn@doi [A\&A]
  {10.1051/0004-6361/201731518}, 608, A57

\bibitem[\protect\citeauthoryear{{Vigna-G{\'o}mez} et~al.,}{{Vigna-G{\'o}mez}
  et~al.}{2018}]{vigna-gomez2018}
{Vigna-G{\'o}mez} A.,  et~al., 2018, \mn@doi [Monthly Notices of the Royal
  Astronomical Society] {10.1093/mnras/sty2463}, 481, 4009

\bibitem[\protect\citeauthoryear{Willcox, Mandel, Thrane, Deller, Stevenson  \&
  {Vigna-G{\'o}mez}}{Willcox et~al.}{2021}]{willcox2021}
Willcox R.,  Mandel I.,  Thrane E.,  Deller A.,  Stevenson S.,
  {Vigna-G{\'o}mez} A.,  2021, arXiv:2107.04251 [astro-ph]

\bibitem[\protect\citeauthoryear{{Willems}, {Henninger}, {Levin}, {Ivanova},
  {Kalogera}, {McGhee}, {Timmes}  \& {Fryer}}{{Willems}
  et~al.}{2005}]{willems2005}
{Willems} B.,  {Henninger} M.,  {Levin} T.,  {Ivanova} N.,  {Kalogera} V.,
  {McGhee} K.,  {Timmes} F.~X.,   {Fryer} C.~L.,  2005, \mn@doi [\apj]
  {10.1086/429557}, \href
  {https://ui.adsabs.harvard.edu/abs/2005ApJ...625..324W} {625, 324}

\bibitem[\protect\citeauthoryear{{Woosley}}{{Woosley}}{1993}]{woosley1993}
{Woosley} S.~E.,  1993, \mn@doi [\apj] {10.1086/172359}, \href
  {https://ui.adsabs.harvard.edu/abs/1993ApJ...405..273W} {405, 273}

\bibitem[\protect\citeauthoryear{{Woosley} \& {Heger}}{{Woosley} \&
  {Heger}}{2012}]{woosley2012}
{Woosley} S.~E.,  {Heger} A.,  2012, \mn@doi [\apj]
  {10.1088/0004-637X/752/1/32}, \href
  {https://ui.adsabs.harvard.edu/abs/2012ApJ...752...32W} {752, 32}

\bibitem[\protect\citeauthoryear{{Xiao}, {Galbany}, {Eldridge}  \&
  {Stanway}}{{Xiao} et~al.}{2019}]{xiao19}
{Xiao} L.,  {Galbany} L.,  {Eldridge} J.~J.,   {Stanway} E.~R.,  2019, \mn@doi
  [\mnras] {10.1093/mnras/sty2557}, \href
  {https://ui.adsabs.harvard.edu/abs/2019MNRAS.482..384X} {482, 384}

\bibitem[\protect\citeauthoryear{Yao et~al.,}{Yao et~al.}{2020}]{yao2020}
Yao Y.,  et~al., 2020, \mn@doi [ApJ] {10.3847/1538-4357/abaa3d}, 900, 46

\makeatother
\end{thebibliography}



\bsp	
\label{lastpage}
\end{document}